\documentclass[sigplan,10pt,nonacm,dvipsnames,balance=false]{acmart}
\setcopyright{none}
\date{}

\usepackage{subcaption}
\usepackage{tikz}
\usetikzlibrary{fit,positioning,arrows.meta}
\usepackage{listings}
\usepackage{xspace}
\usepackage{hyphenat}
\usepackage{tabularx}

\usepackage{amssymb}
\usepackage[ruled,linesnumbered,shortend]{algorithm2e}

\SetAlFnt{\normalfont\normalsize}
\SetAlCapFnt{\normalsize}
\SetAlCapNameFnt{\normalsize}
\SetAlgoNlRelativeSize{0}
\SetAlgoCaptionSeparator{:}
\usepackage{enumitem}

\usepackage{xurl}

\hypersetup{hidelinks}

\providecommand{\showeprint}[2][]{%
  \href{https://arxiv.org/abs/#2}{arXiv:\nolinkurl{#2}}}

\AtBeginDocument{\let\footnotesize\normalsize}

\newcommand{\sysname}{CCAMoE\xspace}

\title[CCAMoE: Offloaded MoE Inference]{CCAMoE: Cache-Conditioned Adaptive Inference for Offloaded MoE}

\newcommand{\ustcaffiliation}{%
  \affiliation{%
    \institution{University of Science and Technology of China}
    \city{Hefei}
    \country{China}}}

\author{Xuan Zhang}
\ustcaffiliation
\email{xuanzhang@mail.ustc.edu.cn}

\author{Haisheng Tan}
\ustcaffiliation
\email{hstan@ustc.edu.cn}

\author{Han Tian}
\ustcaffiliation
\email{henrytian@ustc.edu.cn}

\author{Jialiang Huang}
\ustcaffiliation
\email{jialianghuang@mail.ustc.edu.cn}

\author{Yuteng Wang}
\ustcaffiliation
\email{wangyuteng@mail.ustc.edu.cn}

\renewcommand{\shortauthors}{Zhang et al.}

\begin{document}
\begin{abstract}
Expert offloading enables Mixture-of-Experts (MoE) inference on memory-constrained GPUs, but expert transfers and CPU execution limit decoding performance. Speculative decoding can mitigate these costs, yet separate draft weights consume scarce GPU memory and longer drafts increase verification demand. In this paper, we present CCAMoE, a unified runtime that coordinates adaptive speculative decoding with concurrent expert-cache updates. Its design has three key features: 1) Cache-only drafting, which reuses resident target experts to generate proposals and routing hints without separate draft weights; 2) Draft-guided cache updates, which anticipate expert demand and overlap transfers with computation to support subsequent drafting and verification; and 3) Cost-aware draft-length control, which balances expected acceptance against execution cost. Together, these mechanisms coordinate proposal generation and cache preparation through a shared evolving cache, improving proposal quality while reducing verification cost. Across various MoE models, GPU platforms, and workloads, CCAMoE consistently achieves the lowest decoding latency, with up to 2.33$\times$ speedup over state-of-the-art baselines.

\end{abstract}


\maketitle

\section{Introduction}
\label{sec:intro}

Local deployment of large language models on consumer-grade GPUs such as
the RTX~4090 is increasingly common, driven by demands for data privacy
and model customization~\cite{powerinfer,ktransformers}. These on-device
scenarios typically involve a single user with low-concurrency,
latency-sensitive interactions. Mixture-of-Experts (MoE) models are a
promising fit: they activate only a small parameter subset per
token~\cite{shazeer2017moe,switchTransformer,mixtral,deepseekv3}, reducing
compute on consumer hardware. Yet their storage scales with the full model
size---Qwen3-30B-A3B~\cite{qwen3} activates 3.3\,B of 30.5\,B parameters
per token, but its 16-bit weights require 61\,GB, far exceeding typical
GPU memory. Expert offloading bridges this gap by keeping most experts in
host memory and caching a subset on the GPU~\cite{mixtraloffloading,moeinfinity}.

\begin{figure}[t]
  \centering
  \includegraphics[width=\columnwidth]{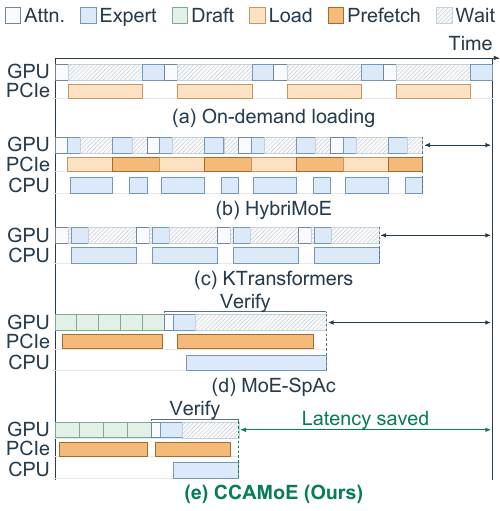}
\caption{Schematic four-token decoding across five methods.
Timelines show full passes, not individual layers.}
  \Description{Schematic comparison of decoding four tokens across five methods.
  Timelines aggregate one full inference pass, not layer-by-layer execution.}
  \label{fig:method-comparison}
\end{figure}

The latency bottleneck in expert-offloaded MoE inference lies in where activated experts reside---on the GPU or in host memory. Figure~\ref{fig:method-comparison}(a--d) illustrates

several representative existing approaches. On-demand loading (a) transfers each activated expert
to the GPU when requested, stalling related computation during transfer.
KTransformers (c) optimizes CPU execution speed and uses static scheduling
with CUDA graphs to reduce kernel launch overhead. However, its static design
forfeits runtime optimization opportunities and depends heavily on CPU
performance, which can be limited on consumer devices. HybriMoE (b)~\cite{zhong2025hybrimoe}
coordinates expert prefetching, CPU computation, and loading, yet the
limited overlap opportunity from autoregressive decoding still leaves
on-demand loading and CPU execution as bottlenecks. Prefetching is a
promising way to mitigate these bottlenecks, but requires both early knowledge of expert demand and sufficient
computation to overlap with weight transfers.

Speculative decoding can provide both. A drafter proposes several
tokens for the target to verify together~\cite{specDecoding,specSampling}.
This reveals a likely continuation and creates a compute window for
expert preparation.
Recent systems exploit these opportunities to forecast expert demand
and coordinate CPU--GPU execution~\cite{moespeq,moespac}
(Figure~\ref{fig:method-comparison}(d)). However, a separate drafter
occupies memory that could otherwise hold target experts. Moreover,
different proposals can activate different experts, so batched
verification must access their combined expert set. This can incur
expert transfers or CPU computation even for proposals that are
ultimately rejected, potentially offsetting the gains from
speculation.

These limitations motivate us to reconsider the role of the GPU expert
cache. Our design builds on two insights.
First, resident target weights can support both approximate drafting
and exact target execution, allowing the same memory allocation to
serve both stages~\cite{specmoe,elmoe3d,ssmoe}. Second, speculation and cache
preparation can reinforce each other: drafting provides early hints
about expert demand and time to fetch them, while cache
updates improve expert availability for later drafting and
verification. This interaction can improve proposal quality while
reducing verification cost.

Therefore, we ask: \emph{Can we build a unified runtime in which
speculation and expert-cache updates reinforce each other to accelerate
offloaded MoE inference?}

Realizing this interaction poses two coupled challenges.
First, the runtime must translate approximate routing hints that arrive
incrementally into timely cache updates while GPU computation uses the
same cache.  These updates must avoid overwriting weights still in
use, preserve efficient CUDA-graph replay, and respect both the GPU memory
budget and available compute windows. Second, the runtime must decide
how long to speculate as cache updates change proposal quality and
verification cost. Longer drafts can yield more opportunities for accepted
tokens and cache preparation, but also add computation and expand the
expert working set for verification.

To address these challenges, we present \sysname, a unified runtime
that coordinates adaptive speculative decoding with concurrent
expert-cache updates (Figure~\ref{fig:method-comparison}(e)).
Its design has three key features:
\begin{itemize}[leftmargin=*]
  \item \emph{Cache-only drafting.} \sysname generates proposals with
  target experts already cached on the GPU, avoiding separate draft
  weights and on-demand expert loading. The same computation also
  provides routing hints about which experts verification may need
  (\S\ref{sec:design:drafting}).
    \item \emph{Draft-guided cache updates.} Weighted by their
  reliability, these hints guide prefetching; where hints are
  unavailable, the runtime falls back to recent target-routing
  history.  A segmented schedule overlaps budgeted transfers with CUDA-graph
  computation and safely updates shared cache slots without a separate
  expert-weight staging buffer.  The newly cached experts
  then support both later draft steps and verification

  (\S\ref{sec:design:cache}).
  \item \emph{Cost-aware draft-length control.} Because the evolving
  cache continually changes both proposal quality and verification
  cost, \sysname balances expected acceptance against drafting and
  verification costs, aiming to
  reduce time per committed output token (TPOT)
  (\S\ref{sec:design:adaptive}).
\end{itemize}

Together, these features coordinate proposal generation and cache
preparation through a shared evolving cache, improving proposal quality
while reducing verification cost.

We evaluate \sysname against five offloaded MoE inference systems
using Qwen3-30B-A3B and DeepSeek-V2-Lite on consumer platforms (RTX~4090
and RTX~3080). \sysname achieves the lowest TPOT across all 12 combinations
of model, platform, and workload. On the RTX~4090, it reaches
46 tokens/s on Qwen3-30B-A3B and 71 tokens/s on DeepSeek-V2-Lite,
yielding 1.78--2.33$\times$ speedups over state-of-the-art baselines
(\S\ref{sec:eval:e2e}).

\section{Background and Motivation}
\label{sec:background}

\subsection{MoE Inference with Expert Offloading}
\label{sec:bg:moe}

Mixture-of-Experts (MoE) models expand model capacity by replacing
dense feed-forward blocks with pools of expert networks.
For each token, a router selects the top-$k$ experts, whose outputs
are combined using routing
weights~\cite{shazeer2017moe,gshard,switchTransformer,mixtral,deepseekv2}.
Only the selected experts are evaluated, so expert computation scales
with the active parameters rather than the total model size. Weight
storage, however, still scales with all experts, since tokens may activate
different subsets.

Expert offloading addresses this memory demand by placing routed-expert
parameters that do not fit on the GPU in host
memory~\cite{mixtraloffloading,moeinfinity}.
Serving a nonresident expert then requires transferring its weights over
PCIe for GPU computation or executing it on the
CPU~\cite{fiddler,ktransformers}.

This saves GPU memory at the cost of transfer or CPU execution latency:
dependent computation must wait until the required expert outputs are
ready. When these operations outlast the GPU work available to overlap
with them, they stall inference~\cite{klotski,layerscope}.
This bottleneck is especially pronounced during decoding at low
concurrency. Whereas prefill processes many prompt tokens together and
reuses expert weights across them, decoding handles only one new token
per request at a time, leaving little computation to amortize each
weight transfer or hide CPU work~\cite{moelightning,ktransformers}.
Because each step depends on the previous step's output, later
steps cannot simply run ahead to fill these gaps, so sparse activation
alone cannot ensure low latency when expert access is slow.

\subsection{Existing Approaches and Their Limitations}
\label{sec:moti:limitations}

\paragraph{Coordinating CPU--GPU execution.}
Prior systems treat expert placement and movement as an online scheduling
problem: predict which experts upcoming layers will need, prefetch their
weights while the GPU is busy, and resolve residual misses through
on-demand PCIe transfers or CPU-side
execution~\cite{mixtraloffloading,moeinfinity,fiddler,zhong2025hybrimoe,
finemoe,layerscope,moelightning,klotski,cloudlocalmoe}.
However, their benefit is constrained by the short overlap windows of
low-concurrency decoding, which limit how much expert transfer and CPU
work can be hidden~\cite{klotski,moespeq}.
Prediction errors further waste bandwidth and can evict useful cached experts,
while prefetches compete with demand loads for PCIe
bandwidth~\cite{finemoe,layerscope}.

KTransformers~\cite{ktransformers} instead uses largely fixed expert
placement, optimized CPU kernels, and CUDA graphs to reduce execution
and coordination overhead. Keeping experts on the CPU avoids weight
transfers, but forgoes the opportunity to reduce CPU expert
computation by transferring their weights to the GPU.
Its performance therefore depends heavily on CPU execution throughput. On consumer systems with less capable CPUs, this
dependence can make CPU expert execution a more severe
bottleneck~\cite{ktransformers,layerscope}. Our CPU-thread sweep illustrates this dependence: on Qwen3-30B-A3B
with MT-Bench on the RTX~4090 platform, reducing CPU threads
from 16 to 4 increases KTransformers' TPOT from 49.7 to
90.4\,ms/token (Figure~\ref{fig:cpu-threads}).

\paragraph{Cache-aware rerouting.}
Cache-aware rerouting exploits expert redundancy to favor resident experts,
reducing cache misses by changing which experts process each
token~\cite{cacheprior,swapmoe,buddymoe}.
However, controlling accuracy loss requires conservative rerouting,
which limits how much CPU expert computation can be bypassed and thus
restricts the achievable speedup~\cite{buddymoe}. Even conservative
rerouting still commits approximate hidden states, KV cache entries,
and generated tokens without exact verification or rollback. Returning to the original router does not restore this altered history.
These deviations can accumulate during longer generation,
amplifying accuracy loss.

\paragraph{Scheduling around a separate speculative drafter.}
Speculative decoding~\cite{specDecoding,specSampling} uses a lightweight draft
model to propose multiple tokens that the target verifies in a single batched pass. Accepted draft tokens advance the sequence by multiple tokens per round, while verification preserves the target model's output distribution.

For memory-bound dense models, batched verification reuses the same
weights across tokens and can add only modest latency over a single
decode step~\cite{specDecoding,specSampling}.
In MoE models, however, different proposals may activate different
experts, so verification must access and execute the union. This
expands the expert working set and makes verification latency depend on
expert reuse as well as draft length, potentially reducing the benefits of speculative
decoding~\cite{moespeq,spmoe,specmoeoff}.
Expert offloading magnifies this cost: additional nonresident experts
require PCIe transfers or CPU computation, including work for tokens
that are ultimately rejected~\cite{spmoe,moespac}.
To mitigate these costs, recent systems combine speculation with expert
prefetching and CPU--GPU scheduling, using the drafting phase to prepare
for verification~\cite{moespeq,spmoe,moespac,specmoeoff}.

However, a separate drafter consumes scarce GPU memory that can otherwise cache target-model
experts~\cite{moespac}. This introduces a size--quality trade-off:
shrinking or quantizing the drafter reduces its footprint but can lower
proposal acceptance~\cite{spmoe,moespeq}.
Even small or low-precision drafters reduce target-expert residency,
potentially increasing CPU computation or transfers, offsetting gains
from accepted drafts.
MoE-SpAc's draft weights alone occupy approximately
4\,GB for Qwen and 9.54\,GB for DeepSeek, accounting for 16.7\%
and 39.8\% of a 24\,GB GPU memory, respectively.
Under tighter memory constraints, MoE-SpAc with an INT4-quantized
DeepSeek-V2-Lite drafter runs out of memory on the 10\,GB RTX~3080
(Figure~\ref{fig:eval:e2e}).

\subsection{Coordinating Speculation and Expert Caching}
\label{sec:moti:observation}

This competition for GPU memory motivates us to reconsider the role of
the resident target weights. We observe that the GPU-resident portion of an offloaded MoE model
could itself serve as a drafter: its resident non-expert parameters and
cached experts can generate approximate proposals without separate
draft weights or on-demand expert loading.
This reuse also suggests a connection between speculation and cache
preparation: the target's router, operating on draft hidden states,
could provide early hints about expert demand, while draft computation
provides time to update the cache. Furthermore, the difference between
the original routes and the cached experts actually used may indicate
how much to trust these proposals and routing hints. These possibilities
lead us to ask whether the cache can support both speculative
progress and its own adaptation.

To examine these possibilities, we run a draft--verify probe on
Qwen3-30B-A3B-Base~\cite{qwen3basemodel} with logical caches holding
16, 32, or 40 of the 128 experts per layer. In each round, the drafter
starts from a verified prefix and proposes 8 tokens using the top-8
resident experts. The target then evaluates these proposals from the
same prefix. Each cache is initialized from recent target-route
frequencies and remains fixed during drafting. Between rounds, we
update it with least-recently-used (LRU) replacement based on committed
target routes. For each cache capacity, we report five rounds per
request, collected after three warm-up rounds.

\begin{figure}[tb]
  \centering
  \includegraphics[width=\columnwidth,trim=0 0bp 0 5bp,clip]{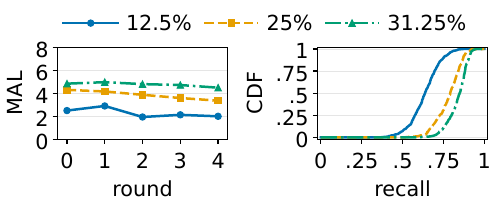}
  
  \par\noindent\makebox[\columnwidth][l]{%
    \hspace*{0.115\columnwidth}%
    \makebox[0.340\columnwidth]{(a)}%
    \hspace*{0.180\columnwidth}%
    \makebox[0.340\columnwidth]{(b)}%
  }
  \caption{Cache-only drafting across expert cache ratios.
  (a)~Mean accepted length (MAL) of 8 proposals.
  (b)~Per-round verifier-route recall CDF.}
  \Description{Two plots show accepted proposal length over measured
  rounds and the distribution of verifier-route recall. Both improve as
  expert residency increases from 12.5 to 31.25 percent.}
  \label{fig:motivation:utility}
\end{figure}

\paragraph{Observation 1: cached experts can draft.}
The cache-only drafter consistently produces usable proposals across
the measured rounds (Figure~\ref{fig:motivation:utility}(a)).
With 31.25\% of experts resident, the mean accepted length (MAL) is
4.75 tokens out of 8 proposals. Even with only 12.5\% resident,
MAL is 2.28 tokens per round.
Each measured round starts only after all LRU replacements from the
previous round's committed routes have completed. When updates
instead overlap drafting, transfer windows and update budgets limit how
many experts can be replaced and when they take effect, so proposal
quality may change within a round.

\paragraph{Observation 2: routing hints predict expert demand.}
During drafting, the shared router scores all experts on approximate
hidden states. We use its top-8 choices before cache constraints as
\emph{routing hints}, including nonresident experts that the drafter
does not execute. Figure~\ref{fig:motivation:utility}(b) shows the
distribution of per-round verifier-route recall, averaged over matched
positions and layers. Median recall ranges from 64\% to 85\% across
the tested cache capacities, supporting hint-guided expert prefetching.

\paragraph{Observation 3: disturbance signals reliability.}
Substituting cached experts for nonresident ones changes the
computation behind both token proposals and routing hints.
We quantify this change by \emph{reroute disturbance} $D$: the fraction
of the original top-8 routing weight assigned to nonresident experts,
averaged over layers for each proposal. A larger $D$ therefore means that more of the intended expert
contribution is replaced. Comparing $D$ with the total-variation
distance between the drafter's and verifier's weighted expert
selections yields a Spearman correlation of 0.94 across all proposed
tokens, suggesting that $D$ can indicate draft--target routing
deviation before verification.

\begin{figure}[tb]
  \centering
  \includegraphics[width=\columnwidth,trim=0 0bp 0 3bp,clip]{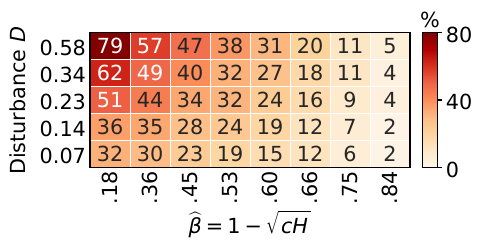}
  \caption{Exact sampling rejection (\%) as a function of the
  acceptance estimate $\widehat{\beta}$ (with $c{=}0.18$) and
  disturbance $D$. }
  \Description{A heat map shows eight of ten fixed-coefficient SVIP acceptance
  estimate bins horizontally and five disturbance quintiles vertically.
  Cells report mean sampling rejection probabilities, which increase
  with disturbance even at similar estimated acceptance.}
  \label{fig:motivation:confidence}
\end{figure}

\begin{figure}[tb]
  \centering
  \includegraphics[width=\columnwidth,trim=0 4bp 0 0bp,clip]{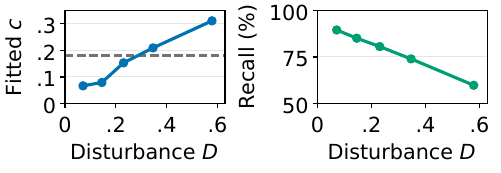}
  \par\noindent\makebox[\columnwidth][l]{%
    \hspace*{0.130\columnwidth}%
    \makebox[0.320\columnwidth]{(a)}%
    \hspace*{0.185\columnwidth}%
    \makebox[0.340\columnwidth]{(b)}%
  }
  \caption{Reliability versus disturbance $D$.
  (a)~Fitted $c$ (dashed: $c=0.18$).
  (b)~Mean verifier-route recall.}
  \Description{Two side-by-side plots share mean disturbance on the
  horizontal axis. The fitted acceptance coefficient increases with
  disturbance, while verifier-route recall decreases. A horizontal
  dashed line marks the fixed coefficient 0.18 in the left panel.}
  \label{fig:motivation:reliability}
\end{figure}

SVIP~\cite{svip} estimates token acceptance using an approximate
lower bound, denoted here by
$\widehat{\beta}=1-\sqrt{cH}$, where $H$ is the natural-log entropy of the
unfiltered draft-token softmax and $c$ is a fixed draft--target mismatch
coefficient.
Figure~\ref{fig:motivation:confidence} uses $c=0.18$ from SVIP's bound
diagnostic. From left to right, rejection generally falls as
$\widehat{\beta}$ increases, showing that the estimate captures the
overall acceptance trend. Within an estimate bin, however, rejection
varies substantially with $D$: at $\widehat{\beta}\approx0.60$, it rises
from 15.4\% to 30.7\% between the lowest and highest disturbance groups.
Thus, a fixed $c$ is insufficient for accurate acceptance estimation
in our cache-only drafter because it misses this variation with $D$.
To capture this variation, we fit $c$ separately in five disturbance
quintiles so that $\widehat{\beta}$ does not exceed exact sampling
acceptance at 90\% of positions in each group.
Figure~\ref{fig:motivation:reliability}(a) shows the fitted coefficient
increasing from 0.07 to 0.31 with $D$: the same empirical coverage
requires a smaller coefficient at low $D$ and a larger one at high $D$,
supporting disturbance-conditioned calibration as the cache evolves.

Disturbance also indicates how much to trust the routing hints from
Observation~2. In Figure~\ref{fig:motivation:reliability}(b), their mean
verifier-route recall falls from 89.5\% in the lowest-disturbance group
to 59.8\% in the highest. Thus, a signal available during drafting could
guide when to stop drafting and how much to trust routing hints
for prefetching.

Recent efforts~\cite{specmoe,elmoe3d,ssmoe} have explored self-speculative decoding with subsets of target experts, avoiding separate draft models. 
Yet SpecMoE and ELMoE-3D keep expert pools fixed during drafting, underutilizing the draft phase as a transfer window for progressive cache preparation. Meanwhile, SS-MoE relies on short-horizon, layer-local prefetching, leaving less opportunity to hide transfers and complicating static CUDA-graph execution.
To address these limitations, we propose \sysname, a unified runtime that
drafts exclusively from resident experts, removing expert-preparation dependencies from the draft critical path, uses live draft hints to trigger concurrent cache updates for subsequent drafting and verification, 
and adapts draft length to expected acceptance and execution cost. This coordination improves proposal quality and reduces verification cost while preserving exact target verification (\S\ref{sec:design}).

\section{Design}
\label{sec:design}

\subsection{Overview}
\label{sec:design:overview}

\begin{figure*}[t!]
  \centering
  
  \includegraphics[pagebox=cropbox,width=\textwidth]{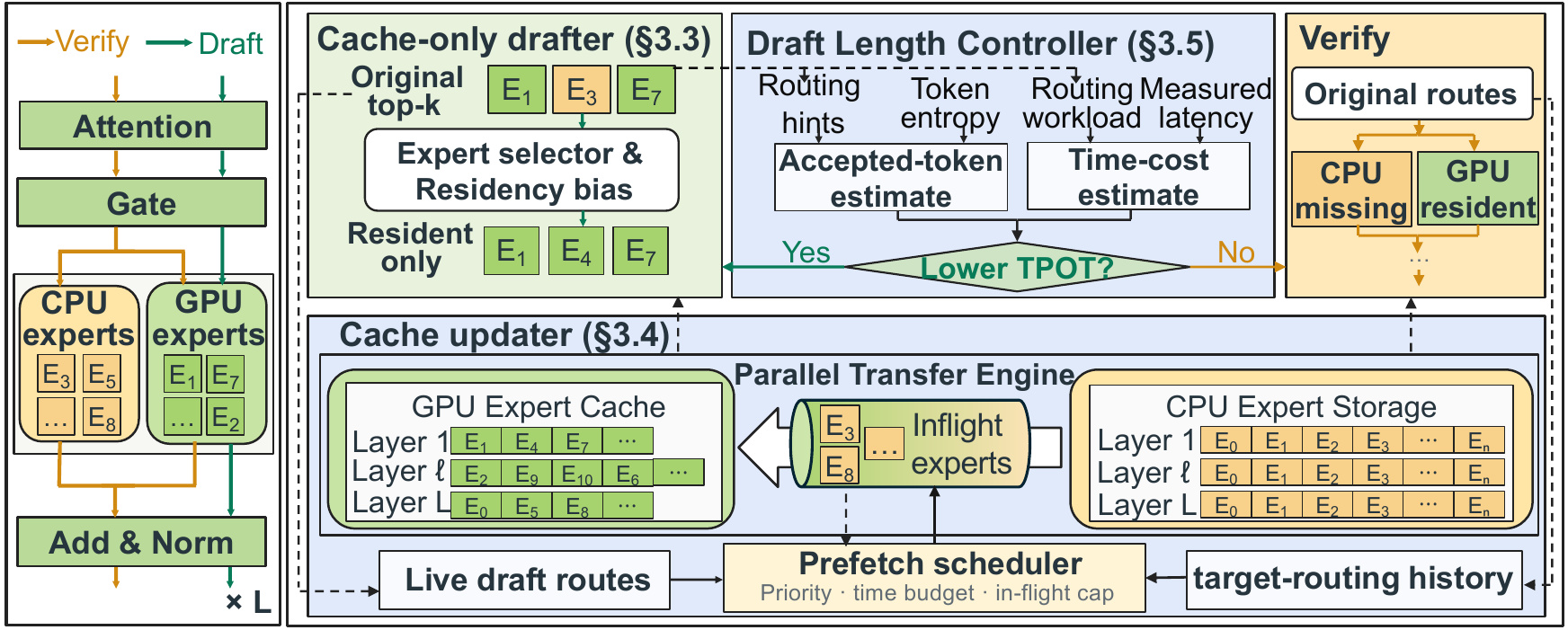}
  \caption{\sysname overview. A shared expert cache supplies drafting
  and GPU expert computation during verification. Routing hints and
  target-routing history guide concurrent cache updates, and acceptance
  and execution-cost estimates determine when to verify. Each box holds
  one expert $E_e$; $e=0,\ldots,n$, and $L$ counts MoE layers.}
  \Description{Cache-conditioned drafting restricts execution to resident
  experts. Adaptive draft-length selection compares expected output with
  execution cost. A cache updater transfers experts from host memory into
  per-layer GPU slots using routing hints and target-routing history.
  Verification executes original routes across CPU and GPU experts.}
  \label{fig:overview}
\end{figure*}

\sysname uses the draft's routing hints and compute window to
actively prepare the shared expert cache. Because updates to this cache
affect both proposal quality and verification cost, the runtime
jointly decides which experts to prefetch and when to stop drafting.

Figure~\ref{fig:overview} follows one speculative round. Starting from a
verified prefix, the \emph{cache-only drafter}
(\S\ref{sec:design:drafting}) reuses resident target weights to generate
proposals without loading missing experts. At each step, the original
router also produces routing hints on draft hidden states, including
choices of nonresident experts. As these hints become available, the
\emph{draft-guided cache updater} (\S\ref{sec:design:cache}) weights
these hints using the disturbance from earlier routing substitutions
and selects experts for prefetching. Recent target-routing
history guides updates where current hints are unavailable. The updater
budgets transfers to fit available compute windows and overlaps them
with computation on other layer segments. Replacements take effect at
segment boundaries before the affected slots are reused, allowing later
draft computation to use the updated cache and preparing experts for
verification.

As drafting proceeds, the \emph{cost-aware draft-length controller}
(\S\ref{sec:design:adaptive}) combines reroute disturbance and token
entropy to estimate proposal acceptance. It uses these estimates and
observed execution costs to compare stopping now with continuing to
longer feasible lengths. Longer drafts provide more time for cache
preparation and may yield more accepted tokens, but also add computation
and expand verification work. The controller continues only when the
estimated TPOT favors doing so.

When drafting stops, the target checks the proposed tokens from the
last verified position using its complete weights and original routing.
Cached experts run on the GPU, while missing experts run on the CPU.
The runtime keeps the accepted prefix and adds one target token to
correct the first rejected proposal or extend a fully accepted draft.
All approximate KV states from drafting are discarded together with the
rejected suffix, leaving only target-computed KV states for the accepted
prefix, so this acceptance and correction procedure preserves the
target's output distribution~\cite{specDecoding,specSampling}.
The target's expert accesses also update the target-routing history for the
next round, which starts from the extended prefix and updated cache.

\subsection{Problem Formulation}
\label{sec:design:formulation}

For a given request, target model, platform, and expert-cache
budget, we seek a policy $\Pi$ that minimizes time per committed
output token. The policy jointly controls proposal generation, cache
updates, draft length, and execution scheduling. Decisions use only available information, respect cache capacity and
execution dependencies, and preserve exact target verification, with
draft lengths constrained to the verifier's supported range and the
remaining output budget.

\paragraph{Expected output.}
Let $r$ index rounds within the request. Given a generated draft of $K$
proposals and their proposal distributions, let $G_r(K)$ be the expected
number of committed tokens. Let $\alpha_{r,i}$ be the probability of
accepting proposal $i$ conditional on acceptance of its preceding
proposals. For a full round without truncation at request termination,
\begin{equation}
G_r(K)=1+\sum_{i=1}^{K}\prod_{j=1}^{i}\alpha_{r,j}.
\label{eq:expected-progress}
\end{equation}
Here $i$ and $j$ index draft positions. Each product is the probability
that the entire prefix through $i$ is accepted; the leading one counts
the correction or bonus token. At termination, $G_r(K)$ counts
only expected committed tokens.

\paragraph{Execution cost.}
Verification evaluates the last verified token and the $K$ proposals
with the preceding target-computed KV context. These inputs determine
which experts are used and how many tokens each expert processes.
Residency when each layer executes splits this work between CPU
and GPU. Let $D_r(K)$ and $V_r(K)$ denote the
drafting and verification times under the policy's cache updates and
execution schedule. Round latency is
$C_r(K)=D_r(K)+V_r(K)$.
Longer drafts increase demand but allow more cache
preparation time, and verification cost reflects both effects.

\paragraph{TPOT objective.}
Let $K_r$ be the draft length in round $r$ and $R$ the number of rounds
until the request ends. Cache updates affect current and later
rounds. We minimize
\begin{equation}
\operatorname{TPOT}(\Pi)=
\frac{\mathbb E_{\Pi}\!\left[\sum_{r=1}^{R}C_r(K_r)\right]}
     {\mathbb E_{\Pi}\!\left[\sum_{r=1}^{R}G_r(K_r)\right]}.
\label{eq:long-run-tpot}
\end{equation}
Here $\mathbb E_\Pi$ averages over sampling and runtime randomness within
this request. By total expectation, the denominator equals the expected
number of committed tokens. The length controller approximates this
request-level objective by comparing estimates of $C_r(K)/G_r(K)$ for
candidate lengths. The following subsections describe the joint policy;
we omit the round index $r$ for brevity.

\subsection{Cache-Only Drafter}
\label{sec:design:drafting}

To avoid waiting for missing experts, \sysname generates proposals
using only GPU-resident experts. We encourage cached substitutes with
a residency bias inspired by Cache-Prior~\cite{cacheprior}, then remove
nonresident candidates before execution. Only drafting is approximate: original-target verification
and correction preserve the output distribution.

At MoE layer $\ell\in\{1,\ldots,L\}$ and draft position $i$, let $g_e$
be expert $e$'s original router logit and $\mathcal C_\ell$ the resident set. Let
$\delta_{\ell,i}\in[0,1]$ be the fraction of original top-$k$ choices
missing from this set, and $\bar H_{\ell,i}\in[0,1]$ the normalized
router entropy over all experts divided by the log of their count.
Entropies use natural logarithms.
We select top-$k$ candidates using the biased logits
\begin{equation}
\widetilde g_e
 =g_e+\gamma(\delta_{\ell,i},\bar H_{\ell,i})\mathbf 1[e\in\mathcal C_\ell],
\label{eq:bias}
\end{equation}
where $\gamma(\delta_{\ell,i},\bar H_{\ell,i})
=\gamma_0\,m(\delta_{\ell,i})\,f(\bar H_{\ell,i})$
and $\gamma_0>0$ sets the base strength. The nonnegative miss gate
$m(\delta_{\ell,i})$ vanishes at zero and grows with misses; the
nonnegative, non-decreasing entropy scale $f(\bar H_{\ell,i})$ strengthens
substitution for diffuse routing.

After selection, the drafter removes nonresident candidates and
renormalizes the original routing weights over the retained set.
If the set is empty, it uses one resident expert with unit weight.
Drafting never loads missing experts on demand or executes them on the CPU.

Each step produces a proposal, tentative KV state, and routing hints
from the original router, including nonresident experts. Earlier
substitutions can alter hidden states, making these hints differ from
verification routes. The updater therefore uses reroute disturbance to
assess hint reliability (\S\ref{sec:design:cache}); the controller uses
the same signal to estimate proposal acceptance
(\S\ref{sec:design:adaptive}).

\subsection{Draft-Guided Cache Updater}
\label{sec:design:cache}

Routing hints provide lookahead for the cache to anticipate expert demand,
but arrive incrementally and may differ from target routes because
expert substitutions in preceding layers alter hidden states.
To use this incomplete and approximate evidence, \sysname falls back
to target-routing history where hints are unavailable and weights
available hints by their reliability. The updater budgets the resulting
transfers against the available compute window to limit exposed transfer
latency. These updates make relevant experts available to later draft
steps and reduce the verifier's CPU workload.

\subsubsection{Expert Cache Organization}
\label{sec:design:cache:organization}
\begin{figure}[tb]
  \centering
  \includegraphics[width=\columnwidth,trim=0 4bp 0 1bp,clip]{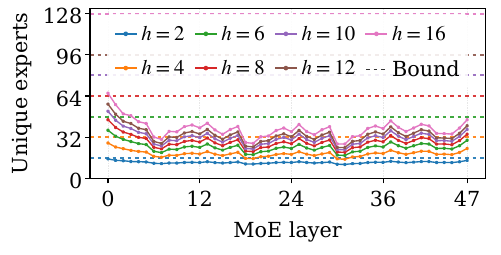}
  \caption{Layer-wise expert demand in Qwen3-30B-A3B, with 48 MoE layers,
  128 experts per layer, and top-8 routing. Curves show mean distinct experts
  in non-overlapping $h{=}2,4,6,8,10,12,16$-token windows; dashed lines
  mark the $\min(8h,128)$ upper bounds.}
  \Description{Layer-wise expert demand retains similar peaks and troughs
  across token-window lengths of 2, 4, 6, 8, 10, 12, and 16, supporting different cache
  capacities across layers.}
  \label{fig:cache:layer-demand}
\end{figure}

To support fused MoE kernels, \sysname fixes the capacity and buffer
layout of each layer's expert cache at initialization.
Figure~\ref{fig:cache:layer-demand} shows that expert working sets differ
across layers, but their relative sizes remain similar across
token-window lengths. We therefore use offline target-routing profiles
to distribute the total slot budget across layers, assigning more slots
to layers with larger working sets. Each slot holds one complete expert;
online updates change its contents and expert-to-slot mapping while keeping buffer
sizes and addresses fixed.

\subsubsection{Parallel Cache Updates}
\label{sec:design:cache:updates}
During drafting and verification, prefetching updates the same expert
cache that computation reads. Concurrent updates must therefore avoid
overwriting weights still in use or letting kernels read incomplete
replacements. CUDA graphs impose an additional constraint: the captured
computation follows a fixed replay sequence, whereas cache scheduling
depends on routing metadata generated at runtime. The scheduler needs
access to routing hints and target routes without inserting data-dependent
host scheduling into the captured computation.

\begin{figure}[t]
  \centering
  \includegraphics[width=\columnwidth]{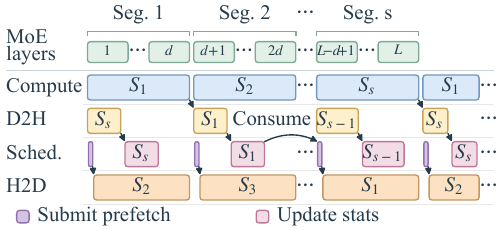}
  \caption{Segmented cache-update pipeline. Computation overlaps metadata
  return and prefetching for the next segment; the final segment prepares
  $S_1$ of the next pass. Segments group $d$ layers (indices $1$--$L$).
  D2H transfers metadata to the host; H2D prefetches experts to the GPU.}
  \Description{A timeline groups MoE layers into segments. Computation
  overlaps the return of routing metadata, cache-update scheduling, and
  expert prefetch for the next segment. Transfers wrap from the final
  segment to the first segment of the following pass.}
  \label{fig:cache-update-pipeline}
\end{figure}

To meet both requirements, \sysname decouples CUDA-graph execution from
dynamic cache scheduling through a segmented pipeline
(Figure~\ref{fig:cache-update-pipeline}). Contiguous
layers form $s$ segments $S_1,\ldots,S_s$, each replayed as a CUDA graph.
While one segment computes, routing metadata from completed segments is
copied asynchronously to the host scheduler. Using the available
evidence, the prefetch scheduler selects experts for the next
segment, and the parallel transfer engine loads them into that
segment's cache
slots, so metadata processing and expert transfers overlap graph
execution. The final segment likewise prepares the first segment of the
following pass.

CUDA events at segment boundaries also enforce safe access to shared
slots: a slot is overwritten only after its old weights are no longer
in use, and the new expert-to-slot mapping is published only after the
transfer completes and before the segment next executes. To keep
transfers hidden by computation, the conservative budget in
\S\ref{sec:design:cache:budget} admits only updates expected to finish
within the current segment's compute window. This ordering permits
direct writes into cache slots during computation on other segments,
avoiding a separate expert-weight staging area and leaving more memory
for resident experts.

\subsubsection{Prefetching and Eviction}
\label{sec:design:cache:policy}
\begin{figure}[tb]
  \centering
  \includegraphics[width=\columnwidth]{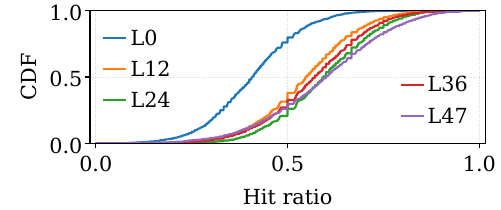}
  \caption{Target-route reuse in Qwen3-30B-A3B: CDF of
  $|\mathcal E_\ell^{-}\cap\mathcal E_\ell^{+}|/|\mathcal E_\ell^{+}|$,
  where $\mathcal E_\ell^{-}$ and $\mathcal E_\ell^{+}$ are expert sets
  in adjacent, non-overlapping 5-token windows (earlier/later).
  Median overlap: $>0.5$ (layers 12, 24, 36, 47),
  $\approx0.4$ (layer 0).}
  \Description{Median adjacent-window expert-set overlap exceeds one half
  for layers 12, 24, 36, and 47, and is about 0.4 for layer 0,
  supporting recent target routes as evidence for upcoming demand.}
  \label{fig:cache:history-reuse}
\end{figure}

The pipeline determines which segment to prepare; the cache policy
selects the experts to bring in and the residents to replace.
At the start of a round, prefetching has no current routing hints to
follow; nearby token windows, however, reuse many of the same experts
(Figure~\ref{fig:cache:history-reuse}), so \sysname retains
target-routing history $\mathcal H$ from prior
verification rounds. For each layer, this history guides prefetching
until its first draft routing record arrives in the current round.
Subsequent updates use that layer's latest draft routing record, while
only target execution updates $\mathcal H$.

Draft hints, however, are not fully reliable: Observation~3 in
\S\ref{sec:moti:observation} shows that greater reroute disturbance
corresponds to lower verifier-route recall
(Figure~\ref{fig:motivation:reliability}(b)), so we weight each hint
by this disturbance. Let $u_{\ell,i}$ and $v_{\ell,i}$ be
the normalized expert-weight distributions from the original top-$k$
routing and actual cache-only execution, respectively, at layer $\ell$
for proposal $i$. Their total-variation distance is
\begin{equation}
\Delta_{\ell,i}
 = 1-\sum_e\min\{u_{\ell,i}(e),v_{\ell,i}(e)\}.
\label{eq:route-deviation}
\end{equation}
Averaging $\Delta_{\ell,i}$ across layers recovers the disturbance $D$
under the motivation probe's routing rule. For prefetching, we retain
the layer-wise values and set a hint's confidence to one minus the mean
disturbance in \emph{preceding} layers:
$\chi_{\ell,i}=1-(\ell-1)^{-1}\sum_{j=1}^{\ell-1}\Delta_{j,i}$
for $\ell>1$, with $\chi_{1,i}=1$ for the first MoE layer.

For each candidate expert $e$, let $R_{\ell,e}$ and $n_{\ell,e}$ denote
its total routing weight and activation count in the selected evidence.
Draft routing records use the latest observation for each layer,
while the history accumulates observations, discounting an expert's
previous totals by a decay factor before adding its new weights and
counts. We compute
the demand score and prefetch priority as
\begin{equation}
\begin{aligned}
M_{\ell,e}&=w_{\mathrm{src}} R_{\ell,e}+\omega n_{\ell,e},\\
\pi_{\ell,e}&=\bar\chi_{\ell,e}M_{\ell,e}-\eta_{\mathrm{age}} a_{\ell,e},
\end{aligned}
\label{eq:cache-priority}
\end{equation}
where $w_{\mathrm{src}}$ weights the source (draft or verification),
and $\omega$ weights counts. A candidate's stored confidence
$\bar\chi_{\ell,e}$ equals $\chi_{\ell,i}$ for its latest draft routing
record at position $i$, averaged across positions for multi-token records;
target history gives $\bar\chi_{\ell,e}=1$. The age $a_{\ell,e}$ counts
runtime inference steps since that observation, with penalty
$\eta_{\mathrm{age}}\geq0$.
We discard observations older than a retention horizon and
consider only positive-priority experts that are neither GPU-resident
nor scheduled for transfer.

Candidates are considered in decreasing priority. Each admission uses
an empty slot when available and otherwise evicts an LRU
resident based on \emph{target} accesses. Newly admitted experts are
protected until verification reaches their layers, giving prefetches a
chance to be used; this protection is relaxed only when no other slot
is available, and slots with unfinished updates remain excluded. Target
experts still absent at verification fall back to CPU execution.

\subsubsection{Transfer Budget}
\label{sec:design:cache:budget}
Prefetching more experts can save the verifier from slow CPU expert
computation, but excessive transfers can stall GPU computation and offset these savings.

The updater therefore limits transfers to
a conservative compute window $W$, estimated separately for drafting
and each verification shape. Automatic startup warm-up calibrates the
per-expert update cost $\widehat\tau$, including transfer time and
an allowance for publishing the replacement. These estimates give the
nominal admission budget
$B=\min\{B_{\max},\lfloor W/\widehat\tau\rfloor\}$,
where $B_{\max}$ caps admissions and $W$ is a conservative fraction of
the observed segment time. Each update requires a replaceable slot,
and a global cap on unfinished transfers prevents work from accumulating
across segments.

During verification, the updater also accounts for elapsed scheduling
time and queued transfers before each admission. It stops submitting
when the current segment finishes or the next update no longer fits
within the window. Online timings refine the cost estimates and adjust
the budget: update batches that exceed the compute window reduce future
budgets, while several consecutive batches that fully use the budget
and finish on time allow a gradual increase.

Algorithm~\ref{alg:cache-updater} assigns at most $B$ updates to the next
segment using these priorities and eviction rules.

\begin{algorithm}[tbp]
\SetAlgoLined
\caption{Budgeted expert cache updates}
\label{alg:cache-updater}
\KwData{Eligible candidates $\mathcal Q$ for the next segment;
scores $\pi_{\ell,e}$ from Eq.~\ref{eq:cache-priority}; budget $B$;
layer-$\ell$ cache slots indexed by $b$; available slots
$\mathcal F_\ell$ (no unfinished updates), protected slots
$\mathcal P_\ell$, last target access $t_\ell(b)$ of slot~$b$'s resident}
\KwResult{At most $B$ assignments $\mathcal U=\{(\ell,b,e)\}$}
$\mathcal U\gets\varnothing$\;
\ForEach{$(\ell,e)\in\mathcal Q$ in decreasing $\pi_{\ell,e}$}{
  \If{$|\mathcal U|=B$}{
    \textbf{break}\;
  }
  \If{$\mathcal F_\ell=\varnothing$}{
    \textbf{continue}\;
  }
  \eIf{an empty slot exists in $\mathcal F_\ell$}{
    $b\gets$ an empty slot in $\mathcal F_\ell$\;
  }{
    $\mathcal V\gets\mathcal F_\ell\setminus\mathcal P_\ell$\;
    \If{$\mathcal V=\varnothing$}{
      $\mathcal V\gets\mathcal F_\ell$\tcp*{relax protection}
    }
    $b\gets\arg\min_{b'\in\mathcal V}t_\ell(b')$\;
  }
  $\mathcal U\gets\mathcal U\cup\{(\ell,b,e)\}$;\quad
  $\mathcal F_\ell\gets\mathcal F_\ell\setminus\{b\}$\;
}
\Return{$\mathcal U$}\;
\end{algorithm}

\subsection{Cost-Aware Draft-Length Controller}
\label{sec:design:adaptive}

A longer draft may yield more committed tokens, but its benefit depends on the drafting and verification time required to obtain them. \sysname therefore compares candidate lengths by the ratio of estimated execution time to expected committed tokens, using this round-level approximation to the TPOT objective in \S\ref{sec:design:formulation} to decide when to stop drafting.

\begingroup
\setlength{\emergencystretch}{2em}
\paragraph{Estimating useful progress.}
Token entropy alone misses the errors introduced by expert
substitution (\S\ref{sec:moti:observation}). Building on SVIP's entropy-based
estimate~\cite{svip}, we scale its mismatch coefficient by reroute
disturbance. For proposal $i$, the base estimate is
\par\endgroup
\begin{equation}
\widehat\alpha_i^{\mathrm{base}}
 =\max\!\left\{0,\,1-\sqrt{\lambda\Delta_i H_i}\right\},
\label{eq:guarded-alpha}
\end{equation}
where $H_i=-\sum_x q_i(x)\log q_i(x)$ is draft-token entropy over vocabulary
tokens $x$. Here $q_i$ is the sampling distribution after temperature and
filtering, or the unfiltered unit-temperature softmax for greedy decoding.
The mean disturbance is $\Delta_i=L^{-1}\sum_{\ell=1}^{L}\Delta_{\ell,i}$
(Eq.~\ref{eq:route-deviation}), and $\lambda>0$ controls sensitivity,
generalizing SVIP's fixed mismatch coefficient $c$.
Let $\widehat\alpha_i$ denote the estimate after an additional discount
when the GPU cache covers little of the original routing weight.

At current length $K$, the controller retains generated proposals'
$\widehat\alpha_i$ values and forecasts each future position with their recent
minimum. Substitution into Eq.~\ref{eq:expected-progress} defines
$\widehat G_K(K')$, the committed-output estimate for candidate length
$K'\geq K$ based on information at length $K$.

\paragraph{Choosing when to verify.}
Verification cost can be nonmonotonic because GPU execution shapes and
CPU expert batching vary across draft lengths. Looking only one step
ahead can therefore miss a better stopping point. The controller
considers all longer lengths in $\mathcal K$, the set supported by the
verifier and the remaining output budget. Startup calibration and
online timings estimate their total drafting and verification cost
$\widehat C(K')$, with route observations refining the expected CPU
cache-miss work. It selects
\begin{equation}
K^*=\arg\min_{K'\in\mathcal K,\ K'>K}
  \frac{\widehat C(K')}{\widehat G_K(K')}.
\label{eq:lookahead-length}
\end{equation}
Drafting continues only if the candidate's estimated TPOT is no higher
than that of verifying now; the controller also stops to verify when no
longer length or required cost estimate is available. After each
continuation to the next supported length, it updates the estimates and
repeats the comparison.

\section{Implementation}
\label{sec:impl}

We implement \sysname on a nano-vllm~\cite{nanovllm} fork with fused
MoE kernels~\cite{nanovllmfused}, adding about 26K lines of
Python and Triton.
Fused MoE kernels resolve expert identities through device-side mapping
tables, while host workers process the asynchronously copied routing
metadata and schedule transfers; a fused GPU kernel then publishes the
updated mappings at the synchronization points described in
\S\ref{sec:design:cache:updates}. During target verification, resident
experts run through the GPU kernels and nonresident experts execute
through KTransformers' CPUInfer backend~\cite{ktransformers}, whose
outputs return through pinned host buffers and are combined with the GPU
outputs using the original routing weights.

\section{Evaluation}
\label{sec:eval}

We evaluate \sysname's decoding performance across models and platforms
(\S\ref{sec:eval:perf}), analyze the effects of cache capacity, draft
length, and concurrent cache updates (\S\ref{sec:eval:spec}), and
measure the contribution of each component (\S\ref{sec:eval:ablation}).

\begin{figure*}[t]
  \centering
  \includegraphics[width=\textwidth,trim=0 3bp 0 7bp,clip]{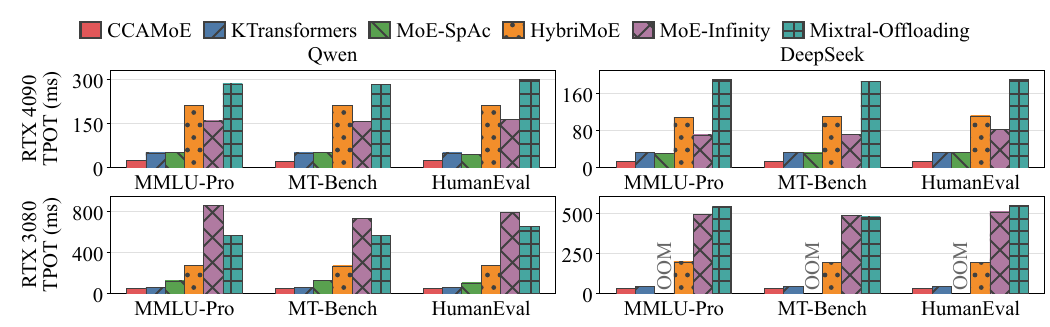}
  \caption{End-to-end decode TPOT (lower is better); OOM marks
  out-of-memory failures.}
  \Description{Four panels compare six systems across the evaluated workloads.
  CCAMoE has the lowest reported TPOT in every model-platform-workload
  combination. MoE-SpAc is out of memory for DeepSeek on the RTX 3080.}
  \label{fig:eval:e2e}
\end{figure*}

\subsection{Experimental Setup}
\label{sec:setup}

\paragraph{Hardware.}
We conduct experiments on the platforms summarized in
Table~\ref{tab:hw}. CPU expert execution uses 16 threads on both
platforms unless otherwise specified.

\begin{table}[htbp]
  \centering
  \caption{Hardware platforms.}
  \label{tab:hw}
  \normalsize
  \begin{tabularx}{\columnwidth}{@{}l>{\centering\arraybackslash}X>{\centering\arraybackslash}X@{}}
    \toprule
     & \textbf{4090 platform} & \textbf{3080 platform} \\
    \midrule
    GPU       & RTX~4090, 24\,GB   & RTX~3080, 10\,GB \\
    Host link & PCIe 4.0$\times$16 & PCIe 3.0$\times$16 \\
    Host CPU  & EPYC~9554          & Xeon Gold 5218R \\
    DRAM      & 1\,TB              & 128\,GB \\
    \bottomrule
  \end{tabularx}
\end{table}

\paragraph{Models.}
We use Qwen3-30B-A3B~\cite{qwen3model} and
DeepSeek-V2-Lite~\cite{deepseekv2lite}.
Table~\ref{tab:models} summarizes their configurations.
All systems run 16-bit weights (BF16/FP16) for these target models.

\begin{table}[htbp]
  \centering
  \caption{MoE model configurations. Expert-weight sizes are approximate
  totals for routed experts in BF16.}
  \label{tab:models}
  \normalsize
  \setlength{\tabcolsep}{1pt}%
  \begin{tabularx}{\columnwidth}{@{}l>{\centering\arraybackslash}X>{\centering\arraybackslash}X@{}}
    \toprule
     & \textbf{Qwen3-30B-A3B} & \textbf{DeepSeek-V2-Lite} \\
    \midrule
    Total / active     & 30.5B / 3.3B & 15.7B / 2.4B \\
    Total / MoE layers    & 48 / 48      & 27 / 26 \\
    Experts / layer       & 128          & 64 + 2 shared \\
    Top-$k$               & 8            & 6 \\
    Expert weights        & 58\,GB      & 30\,GB \\
    \bottomrule
  \end{tabularx}
\end{table}

\paragraph{Workloads.}
We use MT-Bench~\cite{mtbench}, HumanEval~\cite{humaneval},
and the MMLU-Pro validation set~\cite{mmlupro}, covering conversation,
code generation, and knowledge reasoning, respectively. We additionally
use English requests from LongBench-Write~\cite{longwriter} to evaluate
long-generation performance (\S\ref{sec:eval:length}). All experiments use a batch size
of 1, representing a typical local deployment
scenario~\cite{ktransformers}. 

\paragraph{Baselines.}
We compare \sysname with $5$ offloaded MoE inference systems. KTransformers~\cite{ktransformers}
uses CPU--GPU hybrid execution; we disable its Expert Deferral,
which trades accuracy for speed. HybriMoE~\cite{zhong2025hybrimoe}
dynamically schedules expert computation across CPU and GPU.
MoE-Infinity~\cite{moeinfinity} uses activation-aware prefetching.
Mixtral-Offloading~\cite{mixtraloffloading} combines expert caching
with offloading. MoE-SpAc~\cite{moespac} uses speculation to assist
expert scheduling; following its original settings, we pair it with
Qwen3-4B-FP8 as the draft model for Qwen and INT4-quantized
DeepSeek-V2-Lite for DeepSeek.

\subsection{Performance}
\label{sec:eval:perf}

\subsubsection{End-to-End Decoding Performance}
\label{sec:eval:e2e}

Figure~\ref{fig:eval:e2e} compares decode TPOT across models,
platforms, and workloads; \sysname achieves the lowest TPOT in every
evaluated setting. On the RTX~4090, it achieves
$1.78\times$--$2.33\times$ speedups on Qwen and
$2.21\times$--$2.27\times$ on DeepSeek over the strongest baseline
for each workload. For Qwen, this baseline is KTransformers on
MMLU-Pro and MT-Bench, and MoE-SpAc on HumanEval; for DeepSeek,
it is MoE-SpAc throughout. These gains hold whether the strongest
baseline relies on CPU--GPU hybrid execution or on speculation-assisted
expert scheduling. The ablation study in \S\ref{sec:eval:ablation}
highlights the roles of efficient execution and cache preparation:
eager drafting alone increases TPOT, while CUDA-graph
replay and draft-guided cache updates substantially reduce it.
The execution profiles in \S\ref{sec:eval:updates} further show how
cache updates increase accepted progress and reduce CPU expert work
during verification.

On the RTX~3080, \sysname outperforms
KTransformers, the strongest baseline, by $1.34\times$ on Qwen and
$1.20\times$--$1.24\times$ on DeepSeek, while MoE-SpAc is slower than
KTransformers on Qwen and runs out of memory on DeepSeek. \sysname
still sustains speculative execution in this setting because it drafts
from resident target experts within the existing expert cache. The smaller
speedups are consistent with the memory-budget results in
\S\ref{sec:eval:cache}, where limiting expert residency reduces the
benefit of the shared cache. 

\begin{figure}[tbp]
  \centering
  \includegraphics[width=\columnwidth,trim=0 0bp 0 6bp,clip]{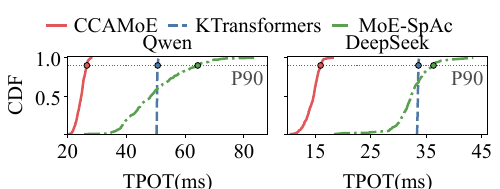}
  \caption{Per-request TPOT distribution (RTX 4090).}
  \Description{Two panels show TPOT distributions for CCAMoE,
  KTransformers, and MoE-SpAc. CCAMoE has lower median and tail
  latency on both models.}
  \label{fig:tpot-cdf}
\end{figure}

Figure~\ref{fig:tpot-cdf} shows the per-request TPOT distributions
on the RTX~4090, pooling traces across workloads. On Qwen and DeepSeek,
\sysname's P90 is 26.6 and 16.0\,ms/token, respectively, below both
KTransformers' and MoE-SpAc's medians for each model.
Although KTransformers shows a steeper CDF, indicating less variation
across requests, its distribution remains at higher TPOT, whereas
MoE-SpAc spans a broader latency range, particularly on Qwen.
\sysname thus lowers both average and upper-tail TPOT.

\subsubsection{Impact of GPU Memory Budget}
\label{sec:eval:cache}

\begin{figure}[tbp]
  \centering
  \includegraphics[width=\columnwidth]{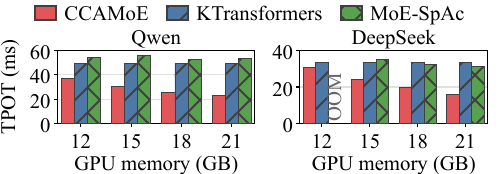}
  \caption{TPOT versus GPU memory (RTX 4090).}
  \Description{Two panels compare CCAMoE, KTransformers, and
  MoE-SpAc across GPU memory budgets. CCAMoE has the lowest
  TPOT throughout. MoE-SpAc is out of memory for DeepSeek at 12 GB.}
  \label{fig:hbm-budget}
\end{figure}

To isolate the effect of GPU memory capacity, we vary the memory
budget from 12 to 21\,GB on the same RTX~4090 using MT-Bench.
Additional memory expands the shared expert cache for drafting and
verification, widening \sysname's advantage
(Figure~\ref{fig:hbm-budget}). At 21\,GB, speedups over
the fastest baseline reach $2.17\times$ on Qwen and $1.94\times$
on DeepSeek. At tight budgets, fewer experts fit in the shared cache,
limiting draft acceptance and leaving more verification work on the
CPU, so speedups are smaller. \S\ref{sec:eval:acceptance}
examines how expert residency affects draft acceptance.
Nevertheless, even at 12\,GB, \sysname outperforms KTransformers,
the fastest feasible baseline for both models, by $1.33\times$
and $1.09\times$, respectively. MoE-SpAc runs out of memory on
DeepSeek at this budget, while \sysname drafts from resident
target experts without separate draft weights.

\subsubsection{Impact of CPU Thread Count}
\label{sec:eval:threads}

\begin{figure}[tbp]
  \centering
  \includegraphics[width=\columnwidth]{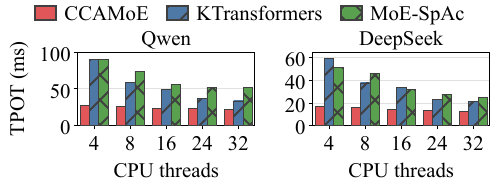}
  \caption{TPOT vs. CPU threads (MT-Bench, RTX 4090).}
  \Description{Two panels compare CCAMoE, KTransformers, and
  MoE-SpAc with 4, 8, 16, 24, and 32 CPU threads. CCAMoE has the
  lowest TPOT at every thread count on both models.}
  \label{fig:cpu-threads}
\end{figure}

We vary the number of CPU threads from 4 to 32 on the RTX~4090
using MT-Bench to evaluate performance under different CPU
configurations. As shown in Figure~\ref{fig:cpu-threads}, \sysname
consistently outperforms KTransformers and MoE-SpAc on both models
at every thread count. Relative to the faster baseline at each
thread count, it achieves $1.54\times$--$3.24\times$ speedups on
Qwen and $1.63\times$--$3.05\times$ on DeepSeek. Both baselines
improve as more CPU threads become available, but \sysname
maintains a substantial performance advantage throughout the
range. Even with few threads, \sysname still benefits from
reduced CPU expert computation: drafting reuses GPU-resident experts, while
concurrent cache updates allow more expert computation to run on
the GPU during verification. In addition, each verification pass
processes multiple proposals, amortizing verification cost over
the accepted output tokens.

\subsubsection{Impact of Output Length}
\label{sec:eval:length}

\begin{figure}[tbp]
  \centering
  \includegraphics[width=\columnwidth,trim=0 2bp 0 0bp,clip]{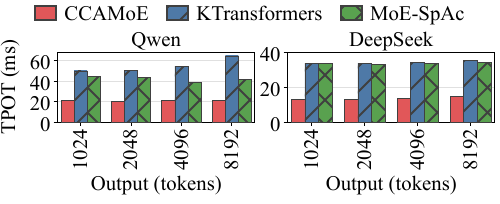}
  \caption{TPOT versus output budget (RTX 4090).}
  \Description{Two panels compare output budgets from 1024 to 8192
  tokens. CCAMoE retains the lowest TPOT in both models, with
  modest variation as the budget increases.}
  \label{fig:output-length}
\end{figure}

Figure~\ref{fig:output-length} shows decoding performance on
LongBench-Write as the output budget increases from 1,024 to 8,192
tokens. \sysname maintains the lowest TPOT across the range, and
its latency grows only marginally with output length---by 3.3\% on
Qwen and 11.8\% on DeepSeek. The baselines are also relatively
flat, except KTransformers on Qwen, which rises more steeply.
\sysname thus retains its advantage even as longer outputs
accumulate a growing KV state.

\subsection{Mechanism Analysis}
\label{sec:eval:spec}

We next examine how cache capacity, draft length, and concurrent
expert-cache updates affect decoding performance.

\subsubsection{Speculative Progress from the Shared Cache}
\label{sec:eval:acceptance}

\begin{figure}[tbp]
  \centering
  \includegraphics[width=\columnwidth]{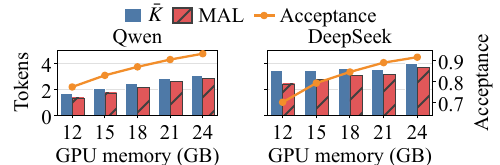}
  \caption{Mean drafted ($\bar K$) and accepted (MAL) tokens per
  round, together with the token acceptance rate, versus GPU memory
  (MT-Bench, RTX 4090).}
  \Description{As memory increases, Qwen's draft length, accepted
  length, and acceptance rate all rise. DeepSeek's accepted length
  and acceptance rate rise while mean draft length rises only
  modestly.}
  \label{fig:mech_accept}
\end{figure}

Figure~\ref{fig:mech_accept} shows how GPU memory capacity affects
speculative progress. On Qwen, increasing the memory budget from
12 to 24\,GB raises token acceptance from 77.5\% to 93.3\% and
MAL from 1.33 to 2.83. Mean draft length also increases, so the
additional progress comes from both longer drafts and a higher
fraction of accepted proposals.
On DeepSeek, token acceptance increases from 70.2\% to 91.6\% and
MAL from 2.40 to 3.64 over the same memory range. Mean draft length
changes more modestly, with improved acceptance accounting for most
of the gain. Notably, acceptance stays above 70\% on both models
even at 12\,GB, evidencing the shared cache's effectiveness under
tight capacity. Enlarging the cache further increases the accepted
progress per verification round on both models, helping explain the
TPOT reductions in \S\ref{sec:eval:cache}.

\subsubsection{Adapting Draft Length to the Memory Budget}
\label{sec:eval:adaptive}

\begin{figure}[tbp]
  \centering
  \includegraphics[width=\columnwidth]{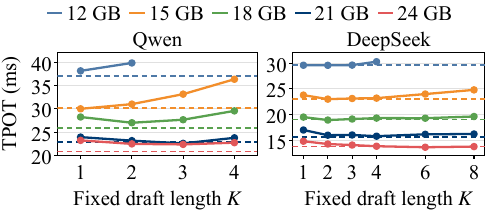}
  \caption{Fixed-$K$ TPOT on the RTX 4090. Dashed lines mark
  adaptive-length runs; out-of-range points are omitted.}
  \Description{Fixed-length curves show that the best tested draft
  length increases with memory capacity. Adaptive runs stay within
  1.25 percent of the best fixed length on DeepSeek and improve on
  it by up to 6.8 percent on Qwen.}
  \label{fig:mech_ksweep}
\end{figure}

To evaluate draft-length selection, we compare fixed lengths with
the cost-aware draft-length controller under different memory budgets
(Figure~\ref{fig:mech_ksweep}). As the budget grows from 12 to
24\,GB, the best fixed length increases from $K{=}1$ to
$K{=}3$ on Qwen and from $K{=}2$ to $K{=}6$ on DeepSeek.
Longer drafts can be costly when memory is limited---at 12\,GB,
increasing $K$ from 1 to 4 raises Qwen's TPOT by 26.0\%---whereas at
24\,GB TPOT varies much less across these lengths. The benefit of
extending a draft thus depends on the cache capacity.

Unlike a fixed $K$, which must be swept offline per budget, the
controller adjusts draft length during execution. On DeepSeek, it
never trails the best fixed length by more than 1.25\% and improves
on it by up to 1.2\% under identical requests and cache capacities.
Qwen's adaptive runs, collected separately, improve on the best
fixed length by 2.8\%--6.8\% at 12, 18, and 24\,GB and stay within
1\% elsewhere.

\subsubsection{Benefits of Concurrent Expert-Cache Updates}
\label{sec:eval:updates}

\begin{figure}[tbp]
  \centering
  \includegraphics[width=\columnwidth,trim=0 2bp 0 4bp,clip]{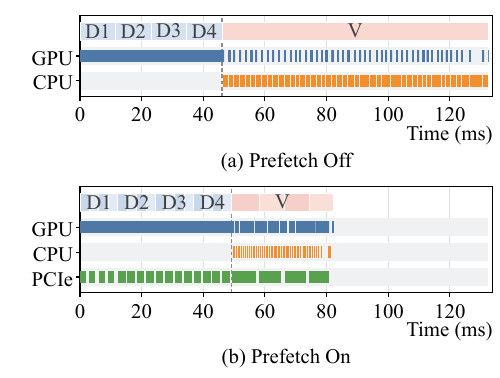}
  \caption{Reconstructed execution profiles with prefetching off (a)
  and on (b) (Qwen, RTX 4090, $K{=}4$). D1--D4 denote drafting steps
  and V denotes verification.}
  \Description{GPU, CPU, and PCIe lanes illustrate execution with
  and without prefetch. The prefetch-enabled view shows transfers
  during drafting and less CPU expert work during verification.
  The reconstructed horizontal axis is not round wall-clock time.}
  \label{fig:pipeline}
\end{figure}

\begin{figure}[tbp]
  \centering
  \includegraphics[width=\columnwidth]{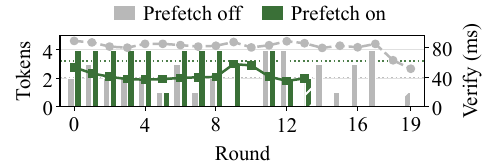}
  \caption{Accepted proposals per round (bars) and verification latency
  (lines) in Figure~\ref{fig:pipeline}. Hatched bars
  are shortened tail rounds; dotted lines average full-$K$ rounds.}
  \Description{The prefetch-enabled profile generally accepts more
  proposals per round and spends less time in verification. Round
  indices refer to each run's own speculative progression.}
  \label{fig:round-acceptance}
\end{figure}

To isolate the effect of concurrent cache updates, we fix the draft
length at $K{=}4$ on Qwen and compare execution with and without
prefetching. Figure~\ref{fig:pipeline} illustrates how updates change
the execution profile: expert transfers overlap drafting, making
more experts available on the GPU and reducing CPU expert work
during verification.
Over full-length rounds after warm-up, prefetching reduces mean
verification latency from 83.4 to 42.9\,ms and increases MAL from
2.06 to 3.23 (Figure~\ref{fig:round-acceptance}). Cache updates
therefore benefit both drafting and
verification: newly resident experts improve subsequent proposals
and reduce the CPU work needed to verify them.

\subsection{Ablation Study}
\label{sec:eval:ablation}

\begin{figure}[tbp]
  \centering
  \includegraphics[width=\columnwidth]{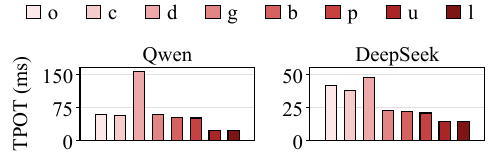}
  \caption{Cumulative ablation on MT-Bench (RTX 4090), adding
  features from left to right: \textup{o}: eager \underline{o}ffloading;
  \textup{c}: fixed uniform expert \underline{c}ache;
  \textup{d}: cache-only \underline{d}rafting;
  \textup{g}: CUDA \underline{g}raphs; \textup{b}: residency \underline{b}ias;
  \textup{p}: \underline{p}rofiled layer capacities;
  \textup{u}: draft-guided cache \underline{u}pdates;
  \textup{l}: cost-aware draft-\underline{l}ength controller.}
  \Description{Eight bars per model share one configuration legend.
  Eager drafting increases TPOT; CUDA graphs and draft-guided cache
  updates provide large reductions. The residency bias reduces TPOT on both
  models. The full system reaches 21.605 ms/token on Qwen and
  14.207 ms/token on DeepSeek.}
  \label{fig:ablation}
\end{figure}

To quantify each component's contribution, we progressively add
\sysname's optimizations to an eager CPU--GPU offloading baseline
that stores all routed experts in CPU memory and executes them on
the CPU, similar to KTransformers~\cite{ktransformers}.
All configurations share the expert kernels and CPU execution path.
Figure~\ref{fig:ablation}
reports TPOT for both models on MT-Bench (RTX~4090); percentage
changes below are relative to the preceding configuration.

A fixed expert cache with uniform per-layer capacities reduces
TPOT by 3.6\% on Qwen and 8.6\% on DeepSeek. Adding cache-only
drafting in eager mode increases TPOT on both models, so reusing
resident experts does not by itself yield a latency benefit.
CUDA-graph replay then reduces TPOT by 62.4\% and 52.2\%,
respectively, as execution overhead dominates cache-only
speculation.
The residency bias (\S\ref{sec:design:drafting}) reduces TPOT by
11.7\% on Qwen and 5.0\% on DeepSeek, and allocating cache
capacity across layers with offline target-routing profiles
(\S\ref{sec:design:cache:organization}) adds a further 2.2\% and
4.5\%.
Draft-guided cache updates provide the largest gain after CUDA
graphs, reducing TPOT by 56.8\% on Qwen and 30.5\% on DeepSeek,
consistent with the profiles in \S\ref{sec:eval:updates}. The
draft-length controller adds a smaller 1.1\% and 0.9\%, with its
cross-budget role covered in \S\ref{sec:eval:adaptive}. The
complete system is $2.66\times$ and $2.91\times$ faster than
eager offloading on Qwen and DeepSeek.

\section{Related Work}
\label{sec:related}

\paragraph{MoE offloading and heterogeneous execution}
Expert caching and prefetching~\cite{moeinfinity,promoe,finemoe}
reduce expert-loading overhead, while CPU--GPU
execution~\cite{fiddler,ktransformers,zhong2025hybrimoe} avoids
loading every required expert onto the GPU.
Multi-batch pipelines~\cite{moelightning,klotski} hide I/O behind
concurrent requests, but such overlap is limited at low concurrency.
\sysname creates overlap within a single request: resident experts
generate proposals without demand loads, while concurrent
updates prepare experts for subsequent computation.

\paragraph{Speculative decoding for MoE}
Prior work studies speculative efficiency~\cite{moesd},
expert prefetching and verification
pipelines~\cite{spmoe,moespeq,moespac,specmoeoff},
self-drafting~\cite{ssmoe,draftexpert}, and cost-aware candidate
selection~\cite{evict,ecospec}.
Separately provisioned draft weights compete with target-expert
residency; resident-expert designs such as SpecMoE~\cite{specmoe} and
ELMoE-3D~\cite{elmoe3d} hold their draft pools fixed until verification.
\sysname advances cache-only speculation by making draft-guided
updates available within the current round, so the same evolving cache
serves drafting and verification without separate draft weights.

\paragraph{Approximate MoE serving}
Compression~\cite{mc_smoe,floe}, adaptive precision~\cite{moeapex},
and cache-aware routing~\cite{cacheprior,buddymoe} reduce serving
costs by changing model computation.
MoE-Spec~\cite{moespec}, AcceptMoE~\cite{acceptmoe}, and
S$^2$-MoE~\cite{s2moe} likewise modify target-expert selection
during speculative verification.
\sysname confines resident-expert substitution to drafting;
original-target verification and speculative correction preserve
the served model's output distribution.

\section{Discussion}
\label{sec:limitations}

\sysname is evaluated at request batch size one, a typical setting for
latency-sensitive local inference. Serving concurrent requests offers
opportunities to share prefetched experts and overlap transfers with
computation across requests, but requires aggregating expert demand and
coordinating per-request draft lengths under shared memory and transfer
budgets. We leave this extension to future work.

\section{Conclusion}
\label{sec:conclusion}

\sysname accelerates offloaded MoE inference by coordinating
cost-aware draft-length control with concurrent expert-cache updates. The same evolving
cache shapes proposal quality and verification cost: resident experts
generate proposals, routing hints guide replacement, and updated
residents serve subsequent drafting and target execution. The
cost-aware draft-length controller adapts to the resulting acceptance--cost
trade-off under a TPOT objective. This coordination requires no separate
draft-model weights and retains exact target verification. Across
Qwen3-30B-A3B and DeepSeek-V2-Lite on an RTX~4090, \sysname delivers
1.78--2.33$\times$ speedups over state-of-the-art baselines.

\label{paper:technical-end}

\hyphenation{Offloading}
\bibliographystyle{ACM-Reference-Format}
\bibliography{references}

\end{document}